\documentclass[twocolumn,showpacs,preprintnumbers,amsmath,amssymb,prl,superscriptaddress]{revtex4-1}
\usepackage{graphicx}
\usepackage{dcolumn}

\usepackage{booktabs}
\usepackage{bm}
\newcolumntype{.}[1]{D{.}{.}{#1}}
\usepackage{float}
\usepackage{color}
\usepackage{hyperref}
\hypersetup{
	breaklinks=true,
	pdfnewwindow=true,      
	colorlinks=true,       
	linkcolor=blue,          
	citecolor=blue,        
	filecolor=blue,      
	urlcolor=blue,          
}

\begin{document}

\title{Stability and metallization of solid oxygen under high pressure}
\author{S. F. Elatresh}
\affiliation{Department of Chemistry and Chemical Biology, Cornell University, Baker Laboratory, Ithaca, NY 14853-1301, USA}

\author{S. A. Bonev}
\email[Electronic address:]{bonev@llnl.gov}
 \affiliation{Lawrence Livermore National Laboratory, Livermore, California 94550, USA}

\begin{abstract}
\noindent
The phase diagram of oxygen is investigated for
pressures from 50 to 130~GPa and temperatures up 1200 K using first
principles theory. A metallic molecular structure with the $P6_3/mmc$ 
symmetry ($\eta^{'}$ phase)  is determined to be
thermodynamically stable in this pressure range at elevated temperatures above the
$\epsilon$(${O_8}$) phase. Long-standing disagreements between theory
and experiment for the stability of
$\epsilon$(${O_8}$), its metallic character, and the transition
pressure to the $\zeta$ oxygen phase are resolved. 
Crucial for obtaining these results are
the inclusion of anharmonic lattice dynamics effects and accurate calculations of exchange
interactions in the presence of thermal disorder. 
\end{abstract}
\date{\today}
\pacs{61.50.Ks,62.50.-p}
\maketitle

%
%
Oxygen has received a great deal of attention because it is a fundamental
 element, one of the most abundant on earth and the
 only one  known with a diatomic molecule that carries a magnetic moment ~\cite{Lundegaard:2006jl}. 
 It has a rich high pressure ($P$) phase diagram with multiple solid phases exhibiting diverse physical properties 
  ~\cite{Freiman20041,doi:10.1107/S0365110X62002248,PhysRevB.37.5364,75a7cc6fac4e4f14a0e5c4aa7fbb2adb,PhysRevB.29.1387,Schiferl:a22092,
 PhysRevB.65.172106,PhysRevLett.93.055502,doi:10.1021/j100366a020,Shimizu:1998, Zhu17012012,PhysRevLett.112.247201}. 
 One of the most interesting among them is the $\epsilon$-phase, which is stable over a large pressure range. 
It has been studied extensively both theoretically
  ~\cite {PhysRevLett.80.5160,PhysRevB.61.6145,PhysRevLett.88.205503} 
 and experimentally by X-ray diffraction~\cite {Johnson:mo0090,PhysRevLett.88.035504} 
 and spectroscopic measurements ~\cite{PhysRevB.54.R15602,PhysRevLett.83.4093,PhysRevB.61.8801,:/content/aip/journal/jcp/86/10/10.1063/1.452547}.
Despite previous theoretical works suggesting that it has $C2/m$ 
symmetry ~\cite {Johnson:mo0090, PhysRevLett.88.035504,PhysRevLett.74.4690} 
and its strong infrared absorption~\cite {Freiman20041} alluding to $O_2$ molecules forming larger units
 \cite {PhysRevLett.83.4093,PhysRevB.61.6145},  only recent experiments have defined the  
exact  structure as $O_{8}$ clusters ($\epsilon$(${O_8}$))
~\cite{Lundegaard:2006jl,PhysRevLett.97.085503,Meng19082008}. Upon compression to 96~GPa, $\epsilon$(${O_8}$) tranforms to the
metallic $\zeta$
phase~\cite{doi:10.1021/j100366a020,PhysRevLett.74.4690}, which is even
superconducting at 0.6~K ~\cite{Shimizu:1998}. 
Experimentally,  Goncharov et al. ~\cite {2011JChPh.135h4512G}
proposed an $\eta^{'}$ phase in the pressure range of 44 to 90 GPa
and at temperatures ($T$) near 1000~K. They suggested it to be an 
isostructure of the $\eta$-O phase previously proposed 
 at low pressures~\cite{PhysRevLett.94.205701,75a7cc6fac4e4f14a0e5c4aa7fbb2adb}. 

Most of the previous theoretical studies have been limited to 0~K and
show a significant disagreement between the calculated and measured  $\epsilon$-$\zeta$
transition pressures,  as well as the structures of the $\epsilon$ and
$\zeta$ phases~\cite{FREIMAN20181}. 
Ma et al. ~\cite{PhysRevB.76.064101}, suggested $C2/m$  as the best candidate for
$\zeta$-$O_{2}$ but it remains a matter of
debate. The  $\epsilon$-$\zeta$  transition occurs at only 35~GPa in their study.
Theoretical confirmation for the $\eta^{'}$ phase has not
been reported yet. Moreover, the experimental evidence for its
stability is not conclusive, and its nature and phase boundaries are unknown.

Here we report results on the phase diagram of solid
oxygen for pressure up to 130~GPa, at both 0~K and finite $T$.  
First, we resolve the existing  inconsistencies
between theory and experiment regarding $\epsilon$
($O_8$) and the $\epsilon$-$\zeta$  transition. Next, we focus on elevated
temperatures where we show that inclusion of anharmonic effects and accurate
exchange energy calculations in the presence of thermal disorder are both
crucial for determining the stability of  $O_2$ phases. 

%
%
            
\begin{figure} [tbh]
 \hspace*{-5mm}
\includegraphics[width=0.46\textwidth, clip]{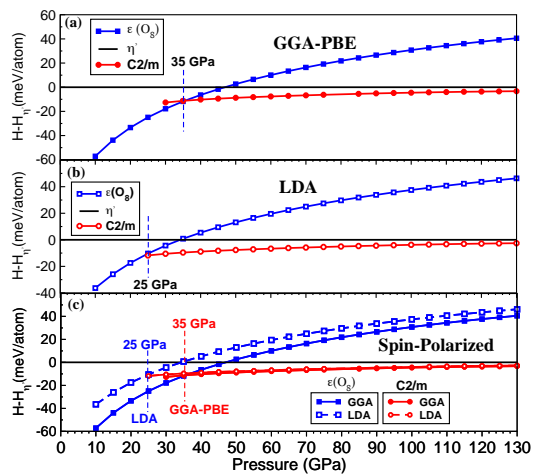}
\caption{ \label{fig: Fig1} (Color online) Enthalpies of the
  $\epsilon$($O_{8}$) and  $\zeta$($C2/m$) oxygen structures relative
  to $\eta^{'}$ computed within the (a) GGA, (b) LDA, and (c)
  spin-polarized LDA and GGA at 0~K. In all cases, $\zeta$($C2/m$) is stable
  in the experimental stability range of $\epsilon$($O_{8}$).}
\end{figure}

      We start by examining oxygen at 0~K in the $P$ range of 10-130~GPa
 and consider the relevant structures $\eta^{'}$ \cite{2011JChPh.135h4512G}, $\epsilon$ ($O_{8}$)  
 \cite{Lundegaard:2006jl},  and $\zeta$ ($C2/m$)  \cite{PhysRevB.76.064101}.
  Density-functional theory calculation (DFT)
  ~\cite{PhysRev.140.A1133} are performed with ABINIT~\cite{2009CoPhC.180.2582G},  using Troullier-Martins Pseudopotentials~\cite{PhysRevB.43.1993} 
  a plane-wave expansion with a 80-Hartree cut off, and  
  { \bf k}-point grids of $16^3$, $4^3$, and $12\times12\times10$ for 
  $\eta^{'}$,  $\epsilon$ ($O_{8}$), and $\zeta$ ($C2/m$) phases,
  respectively, ensuring enthalpies convergence to  better
  than 1~meV/atom. In order to test the effect of the exchange correlation 
 functional approximation of DFT on the relative stability of these structures, calculations  with both the generalized
  gradients approximation  (GGA) and the local density
   approximation (LDA) were performed (Fig.~\ref{fig: Fig1}). The GGA transition between 
    $\epsilon$ ($O_{8}$) and  $\zeta$($C2/m$) is at 35~GPa, which is in agreement 
    with the previous theoretical
    results~\cite{PhysRevB.76.064101}. Note that during structural
    optimization at $P > 50$~GPa, we have restricted the
    occupation of the electronic states of $\epsilon$(${O_8}$) in order to prevent
    its spontaneous transformation to $\zeta$($C2/m$) 
    as reported by Ma et al. ~\cite{PhysRevB.76.064101}. 
Within LDA,  the  $\epsilon$(${O_8}$) -- $\zeta$($C2/m$) transition is 25~GPa and it is 
clear that $\epsilon$(${O_8}$)  is the most sensitive to the choice of exchange correlation functional. 
To examine the significance of spin polarization, we have also performed  spin-polarized calculations within 
GGA  and LDA (Fig.~\ref{fig: Fig1}(c)). 
The results show negligible effects on the enthalpies,
which is in agreement with previous studies reporting 
 that the $O_2$  spin is suppressed at pressures above 10~GPa~\cite{PhysRevLett.94.205701}.

It is well known that the GGA and LDA introduce errors in the ground
state energy calculations that depend on the electronic properties of
the system; the overall tendency is to favor better metals. 
Therefore, the fact that O$_2$ undergoes metallization in the pressure
range of interest and the notable differences among the
electronic properties of the competing structures ($\epsilon$(${O_8}$) is an insulator while $\zeta$($C2/m$) and $\eta^{'}$ 
are metallic; see Supplementary material), raise the question of whether there is a significant
non-cancellation of LDA and GGA errors. 
To examine this, we have carried out hybrid exchange calculations
within the Heyd-Scuseria-Ernzerhof approximation (HSE06) ~\cite{:/content/aip/journal/jcp/124/21/10.1063/1.2204597} as implemented 
 in VASP ~\cite{PhysRevB.47.558}. 
Full structural optimizations within HSE06 were performed using a 1000~eV plane-wave
 cut-off and slightly reduced {\bf k}-point grids:  
  ($12\times12\times4$) for $\eta^{'}$ and ($6\times6\times6$) for
  $\zeta$($C2/m$), ensuring convergence of relative enthalpies
  to better than 2~meV/atom. Additionally, HSE06 electronic
  band structures were calculated on both HSE06 and GGA relaxed
  structures. 

\begin{figure} [t]
 \hspace*{-5mm}
\includegraphics[width=0.46\textwidth, clip]{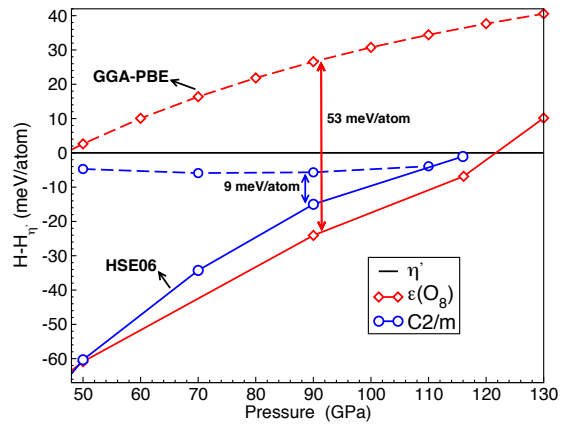}
\caption{ \label{fig: Fig2} (Color online) Enthalpies of $\epsilon$(${O_8}$) 
and  $\zeta$($C2/m$) relative to $\eta^{'}$ oxygen. Dash and solid lines are
GGA-PBE and HSE06 calculations, respectively.  Within HSE06,
$\epsilon$(${O_8}$)  is stable from 50 to 120~GPa, which consistent with the measurement ~\cite{Lundegaard:2006jl}.}
\end{figure}

The results with HSE06 corrections are shown in
Fig.~\ref{fig: Fig2}. The relative enthalpy of  $\epsilon$($O_{8}$) is most
strongly affected. At around 90~GPa, it is lowered by as much as 53 and
44~meV/atom relative to $\eta^{'}$  and $\zeta$($C2/m$),
respectively. These corrections are sufficient to make
$\epsilon$($O_{8}$) the preferred structure in the $P$-$T$ region
where it has been observed experimentally. 
The large effect on $\epsilon$($O_{8}$) is understood by
the fact that (at 0 K) this phase is insulating at $P<107$~GPa
whereupon further compression it metalizes by way of band overlap. The
computed closure of the bandgap at 107~GPa and 0~K is in a good agreement
with the experimental observation of metallization at 
96~GPa~\cite{doi:10.1021/j100366a020,PhysRevLett.74.4690}; indeed, thermal
effects are expected to close the gap at lower pressure. Our analysis
(see Supplementary material) indicates that in order to reproduce the
experimentally observed metallization of  $\epsilon$($O_{8}$)  at
$P> 96$~GPa, it is necessary to perform both structural
optimizations and electronic band structure calculations using hybrid
exchange. To summarize so far, we are able to obtain the
low-$T$ thermodynamic
stability and non-metallic character of oxygen in the $\sim$50--100~GPa
range in
agreement with the experimental observations~\cite{Lundegaard:2006jl}. However, note that our
results do not support the $\epsilon$($O_{8}$)--$\zeta$($C2/m$)
transition at around 107~GPa. Even though the two structures are close in
energy, this suggests that $C2/m$ may not be the correct structure of
the $\zeta$-$O_2$ phase.

\begin{figure} [t]
 \hspace*{-5mm}
\includegraphics[width=0.46\textwidth,clip]{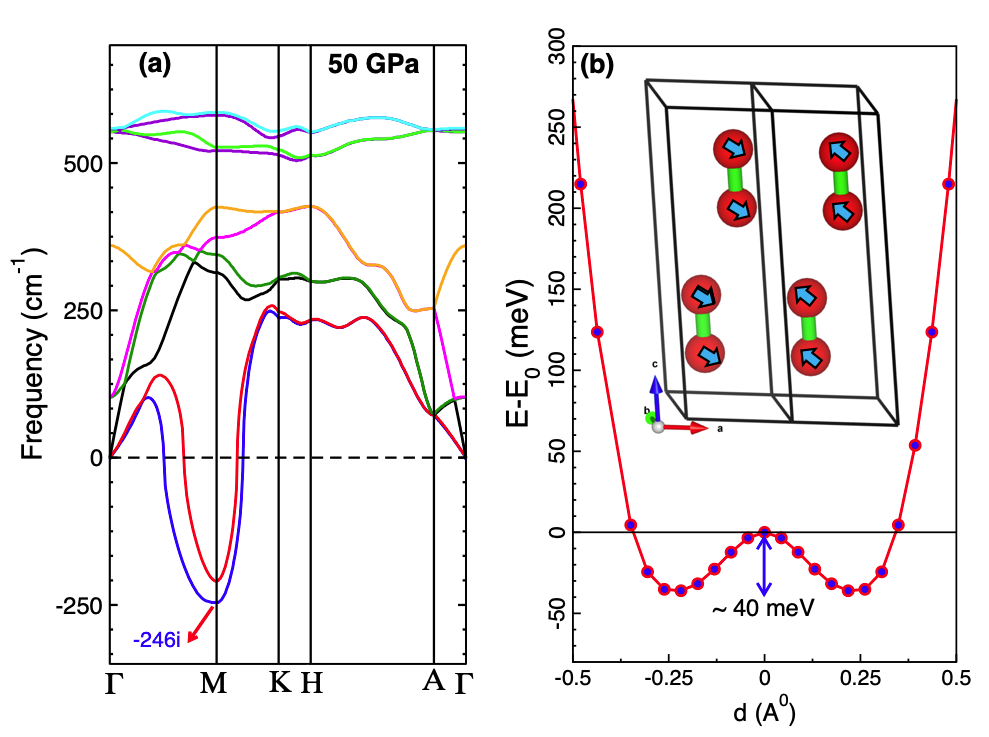}
\caption{\label{fig: band_frozen_phonon} (a) Phonon spectra of 
  $\eta^{'}$ oxygen at 50~GPa computed within the harmonic approximation as
  explained in the text. (b) Change in energy as a function
  of atomic displacement, $d$, from the equilibrium positions for the unstable transverse acoustic mode
  at M. }
\end{figure}

Having established the phase properties of $O_2$ at low $T$, we turn
our attention to its thermodynamic stability at elevated $T$. For this,
we have first computed  the phonon dispersions of the $\eta^{'}$ and
$\epsilon$-$O_{8}$ phases at 50 GPa using Density-Functional
Perturbation Theory (DFPT)~\cite{PhysRevB.55.10337} as employed in 
ABINIT~\cite{2009CoPhC.180.2582G}, with the same convergence
parameters used to calculate their enthalpies. For the $\epsilon$(${O_8}$)
phase, a $4^3$  {\bf q}-point phonon grid for computing
the dynamical matrices was sufficient to achieve  convergence for
Helmholtz free energies and entropy better than 1~meV/atom. 
For the $\eta^{'}$  phase, the dynamical matrices were computed on
$3^2\times5$ ,  $3^2\times9$, $5^3$, and
$7^2\times5$ {\bf q}-point grids, in all cases yielding
imaginary phonon frequencies (see Fig.~\ref{fig:
  band_frozen_phonon}(a)). This could mean that the structure is
mechanically unstable and is probably the reason why it was not found
in previous structure search studies. However, depending on the nature
of the soft phonon modes, it can be stabilized at finite $T$ or even
at 0~K due to quantum zero point motion. 
Hence, we have examined more
closely the phonon modes where the instability is most pronounced using the
frozen phonon method. The results 
in Fig.~\ref{fig: band_frozen_phonon}(b) show that the unstable mode is indeed
a shallow double well potential.
 The appearance of a double well
potential associated with a structural instability is similar to what
has been observed in other elements such as $Ca$~
\cite{PhysRevLett.105.235503}. However, compared to $Ca$, the potential 
barrier in $\eta^{'}$-$O_2$ is much higher. Nevertheless, this mode can be thermally stabilized
at $T$ of around several hundred K and is likely
to contribute significantly to the entropy of the  $\eta^{'}$  phase at
elevated $T$. For determining the thermodynamic stability of
$\eta^{'}$-$O_2$, it is therefore necessary to go beyond the quasi-harmonic approximation.

Gibbs free energies ($G$) at finite $T$ were computed using first 
principle molecular dynamic (FPMD) simulations in the $\sim$49--70~GPa
$P$ range and $T= 500$, 800 and 1200~K, using
finite-$T$ DFT~\cite{PhysRev.140.A1133} 
within PBE-GGA~\cite{PhysRevLett.78.1396} and
VASP~\cite{PhysRevB.47.558}. The simulations were carried out with
300-atom supercells, the $\Gamma$ {\bf k}-point, a 6-electron projector augmented wave
pseudopotential, and a 900 eV plane-wave cut-off 
in the canonical $NVT$ (constant number of particles, $N$, volume
$V$, and $T$) ensemble using Born-Oppenheimer dynamics with a
Nos$\acute{e}$-Hoover thermostat.
For each $V$ and $T$, the system was initially equilibrated within
$2$ ps and then ran
for additional $6$~ps or more using a 0.75~fs ionic time-step.
$G$ was calculated as  $G= E_0 + P_0V + P_{\mathrm{ph}} V +
U_{\mathrm{ph}} - TS $,
where $E_0$ and $P_0$ are the 0~K
DFT energy and pressure, $U_{\mathrm{ph}}$ and $P_{\mathrm{ph}} =-\frac{\partial F_{\mathrm{ph}}}{\partial V }|_{N,T}$ are
the phonon internal energy and pressure, $S$
is the entropy, and $F_{\mathrm{ph}} = U_{\mathrm{ph}} - TS$.
Here $U_{\mathrm{ph}}$ and $S$ are  obtained by  integrating vibration density of states
 (VDOS), which for $\eta^{'}$ are calculated by taking a Fourier transforms
of velocity autocorrelation functions (VACF). 
Although $S$ is calculated using a
harmonic partition function, the VDOS from PFMD represent thermally
renormalizes phonons and capture most of the anharmonic free energy~\cite{PhysRevLett.105.235503}. 
For $\epsilon$($O_{8}$), where the harmonic approximation is
sufficient, VDOS are calculated from DFPT.

\begin{figure} [t]
\includegraphics[width=0.47\textwidth, clip]{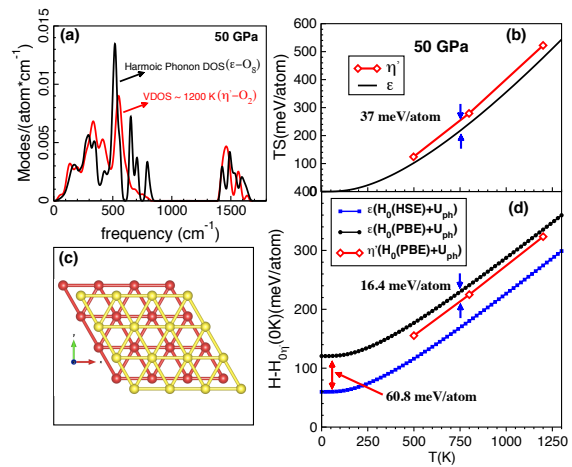}
\caption{ \label{fig: Fig3} (Color online): (a)  Vibration density of
  states (VDOS); (b) $TS$ (temperature times entropy); and (d)
  enthalpies for the $\eta^{'}$ (red) and $\epsilon$($O_{8}$) (black)
  phases relative to the 0~K $\eta^{'}$ structure enthalpy. The relative
  enthalpies are shown with and without HSE06 corrections.
 (c) View of the $\eta^{'}$ phase along the c-axis  (bond direction).}
\end{figure}

A comparison of the VDOS of the $\eta^{'}$ and $\epsilon$(${O_8}$) phases is shown in
Fig.~\ref{fig: Fig3}(a), from which the anharmonicity of the former is
evident. The enthalpies of the two phases as a function of $T$ are
shown in Fig.~\ref{fig: Fig3}(b). Within GGA, the enthalpy of
$\eta^{'}$ remains lower in the entire temperature range of
interest. The lower frequency modes of  $\eta^{'}$-$O_2$ slightly
lower its $H$ relative to that of $\epsilon$(${O_8}$), but the effect is
small and after adding the HSE06 correction computed earlier, the enthalpy
of $\epsilon$(${O_8}$) becomes lower at all temperatures. The soft modes
have much more pronounced effect on the entropy. The $TS$ terms are
potted as a function of temperature in Fig.~\ref{fig: Fig3}(c). As
expected, the  $\eta^{'}$ phase has higher entropy. At 750~K, its $TS$
is 37~meV/atom higher than that of the $\epsilon$(${O_8}$) phase. In the
context of relative stabilities of molecular crystals, this is a
relatively large value. However, it is still not sufficient to overcome
the enthalpy differences between the two phases and to make $\eta^{'}$
preferred at elevated temperatures.

\begin{figure} [tb]
 \hspace*{-5mm}
\includegraphics[width=0.47\textwidth, clip]{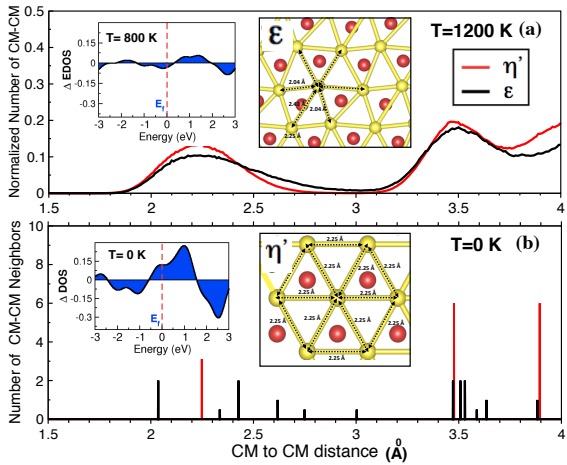}
\caption{ \label{str_analysis} (Color online) Comparison of CM-CM
  distance distributions of the  $\epsilon$(${O_8}$) and
  $\eta^{'}$-$O_2$ structures at 50~GPa and (a) 0~K and (b)
  1200~K. The insets on the right side show the equilibrium atomic
  positions of  $\eta^{'}$ (in (a)) and  $\epsilon$(${O_8}$)
  (in (b)). The insets on the left side show the differences in
  the electronic density of states (EDOS) of the two structures near
  the Fermi level at 0 and 800~K. The difference is much diminished at
  the higher temperature.}
\end{figure}

In order to find a clue that may solve the problem for the finite
temperature stability, we examine the
local structural oder of the $\epsilon$(${O_8}$)  and $\eta^{'}$ phases. The
atomic arrangements in the two structures are shown in the Insets of
Fig.~\ref{str_analysis}. In both cases, the molecules are
arranged in layers, with their bonds perpendicular to the the
layers. Fig.~\ref{str_analysis}(a) shows a distribution of distances
between the molecular center of masses (CM) of the 0~K crystals. The
CM-CM distributions look quite different and it is clear that
$\eta^{'}$ is the more symmetric structure. However, if we examine the
atomic arrangements shown in the Insets of Fig.~\ref{str_analysis}, we
see that locally the $\epsilon$(${O_8}$) phase can be viewed just as a distortion of the
hexagonal molecular arrangements found in the $\eta^{'}$ phase. This
observation suggest that the introduction of thermal disorder may 
bring the average local order of the two structures closer to each
other. CM-CM distance distributions computed from finite-temperature
FPMD trajectories (Fig.~\ref{str_analysis}(b)) show that this is
indeed the case. Namely, the short range orders of the two structures
become similar, on average, in the presence of thermal disorder. As
seen in the Insets in Fig.~\ref{str_analysis}, the difference of the
electronic density of states between the two structures near the Fermi
energy also diminish with temperature. 

The implication of the above observations is that the hybrid exchange
corrections to the energies computed on ideal 0~K crystals may not be
adequate at fine $T$. Indeed, exchange interactions are in principle
short-range. In the HSE06 implementation here, the effective range of
the Hartree-Fock exchange is around 4~A (hence the CM-CM range shown
in Fig.~\ref{str_analysis}). 
Therefore, if the local orders of the two structures become
similar at elevated $T$, then the differences between their HSE06 corrections
are also expected to diminish with temperature. This, in turn, 
will make   $\eta^{'}$-$O_2$ more competitive at finite $T$ compared
to the 0~K case. 
We have therefore performed HSE06 calculations on
atomic configurations taken from the FPMD trajectories at 1200~K. For
each structure,  5-10 equally spaced (in time) configurations were
taken randomly from the trajectories and their energies computed
within GGA-PBE and HSE06 with exactly the same simulation parameters.
We verified that the
fluctuations in the energy differences between HSE06 and GGA-PBE are
negligible, which indicates that the GGA-PBE ensemble is sufficient for this
analysis. 

\begin{figure} [tb]
 \hspace*{-5mm}
\includegraphics[width=0.47\textwidth, clip]{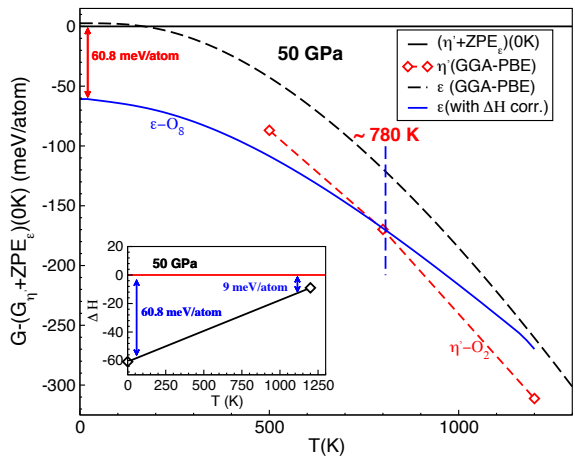}
\caption{ \label{G_vs_T} (Color online) Gibbs free energies
  of $\epsilon$(${O_8}$) and $\eta^{'}$-$O_2$ at 50~GPa as a function
  of temperature, relative to the 0~K result for
  $\eta^{'}$. For comparison, relative energies are computed within 
  both GGA and HSE06. The inset shows the HSE06 corrections to the
  relative  $\epsilon$(${O_8}$)-$\eta^{'}$ enthalpies.}
\end{figure}

 The calculations reveal that at 50~GPa and 1200~K the hybrid exchange
 correction of $\eta^{'}$-$O_2$ relative to  $\epsilon$(${O_8}$) is only
 9~meV/atom -- much smaller than the 0~K value of 60.8~meV/atom. The
 larger entropy of $\eta^{'}$ is now sufficient to compensate for
 this smaller correction and $\eta^{'}$ is thus the preferred phase at
 high $T$ (Fig.~\ref{G_vs_T}). We have
 recomputed the Gibbs free energies of the two structures as a
 function of $T$ by simply interpolating the HSE06 correction
 between the 0 and 1200~K values. The result shows that
 $\eta^{'}$-$O_2$  becomes thermodynamically stable at temperatures
 above about 780~K at 50~GPa, consistent with the experimental observations by
 Goncharov et al.~\cite{2011JChPh.135h4512G}. 

In conclusion, we have shown that $\epsilon$(${O_8}$) is stable
and non-metallic in the $\sim$ 50--100~GPa range, in agreement with
measurements. Computing the structural and electronic
properties of the competing oxygen structures using a beyond GGA exchange
functional is essential for this result. Furthermore, we have determined that the 
experimentally proposed $\eta^{'}$ structure is mechanically unstable at low $T$ within 
a classical ion dynamics treatment due to soft phonon modes. 
However, it
is stabilized at finite $T$ where the thermally renormalized phonons
also contribute to it having a relatively large entropy. In the presence of thermal
disorder, the differences between the local structural order and
electronic properties of $\epsilon$(${O_8}$) and $\eta^{'}$ diminish. The
interplay of all these factors - anharmonicity, exchange effects, and
thermal disorder, results in $\eta^{'}$ becoming the thermodynamically stable phase 
at elevated $T$. At 50~GPa, we predict the transition to take
place at around 780~K. 

%
%

We thank Prof. R. Hoffmann and Dr. V. Askarpour for helpful discussions. This work was supported by NSERC, Acenet and LLNL.  
S.A.B. performed work at LLNL under 
the auspices of the US Department of Energy under contract No. DE-AC52-07NA27344.


%

 \bibliography{ref}
 

 \end{document}